\documentclass[article,submit,moreauthors,galaxies,pdftex,10pt,a4paper]{mdpi} 
\usepackage{subfig}

\firstpage{1} 
\articlenumber{x}
\doinum{10.3390/------}
\pubvolume{xx}
\pubyear{2018}
\copyrightyear{2018}
\history{Received: date; Accepted: date; Published: date}

\Title{Hadronic models of the \textit{Fermi} bubbles: Future perspectives}

\Author{Soebur Razzaque $^{1}$\orcidA{} and Lili Yang $^{2,1}$\orcidB{} }

\AuthorNames{Soebur Razzaque and Lili Yang}

\address{%
$^{1}$ \quad Department of Physics, University of Johannesburg, PO Box 524, Auckland Park 2006, South Africa\\
$^{2}$ \quad Center for Astrophysics and Cosmology, University of Nova Gorica, Vipavska 13, Nova Gorica, Slovenia}

\corres{Correspondence: srazzaque@uj.ac.za; lili.yang@ung.si} 

\firstnote{These authors contributed equally to this work.}
\preto{\abstractkeywords}{\nolinenumbers}
\abstract{The origin of sub-TeV gamma rays detected by \textit{Fermi}-LAT from the \textit{Fermi} bubbles at the Galactic center is still unknown. In a hadronic model, acceleration of protons and/or nuclei and their subsequent interactions with gas in the bubble volume can produce observed gamma rays. Such interactions naturally produce high-energy neutrinos, and a detection of those can discriminate between a hadronic and a leptonic origin of gamma rays. Additional constraints on the \textit{Fermi} bubbles gamma-ray flux in the TeV range from recent HAWC observations restrict hadronic model parameters, which in turn disfavor \textit{Fermi} bubbles as the origin of a large fraction of neutrino events detected by IceCube along the bubble directions. We revisit our hadronic model and discuss future constraints on parameters from observations in very high-energy gamma rays and neutrinos.}

\begin{document}
\section{Introduction}
The \textit{Fermi} bubbles (FB), discovered in the gamma-ray data of the \textit{Fermi} Large Area Telescope (LAT), are two giant structures extending up to 55 degrees ($\sim 9$ kpc) above and below the Galactic center (GC) \cite{Dobler:2009xz, Su:2010qj, Fermi-LAT:2014sfa, Narayanan:2016nzy}. The associated multi-wavelength observations in microwaves (WMAP haze) \cite{Finkbeiner:2003im, Ade:2012nxf}, X-ray \cite{Snowden:1997ze, Tahara:2015mia} and polarized radio waves \cite{Carretti:2013sc} provide comprehensive information for studying the physical origin of the bubbles. Several theoretical models have been proposed to explain the morphology and spectral properties of detected gamma rays, typically classified as hadronic and leptonic models. Both mechanisms can reproduce the observed hard spectrum, sharp edges and uniform emission at latitudes |b| > 10$^\circ$. In the hadronic models, gamma rays are generated by inelastic collisions of accelerated cosmic rays on thermal nuclei in the bubble gas \cite{Crocker:2010dg, Lunardini:2011br, Crocker:2014fla, Fermi-LAT:2014sfa}. While in the leptonic models, inverse Compton scattering of relativistic electrons on optical and UV photons produce the gamma rays \cite{Su:2010qj, Mertsch:2011es, Cheng:2011xd, Fermi-LAT:2014sfa}.  The true origin of gamma rays, whether hadronic or leptonic, has profound implications for the history of star-formation activities in the central region of the Milky Way and/or the activity of the super-massive black hole at the center, as well as particle acceleration.

In hadronic models, TeV to PeV neutrinos, resulting from charged pion/kaon decays, as counterparts of GeV gamma rays, resulting mostly from neutral pion decays, should exist \cite{Crocker:2010dg, Lunardini:2011br}.  Detection of these high energy neutrinos can serve as one of the major discriminator between the hadronic and leptonic models. It was pointed out in the past that a fraction of the high-energy astrophysical neutrinos detected by IceCube \cite{Aartsen:2013jdh} could originate from the FB in hadronic models \cite{Razzaque:2013uoa, Ahlers:2013xia, Lunardini:2013gva, Lunardini:2014wza, Fang:2017vlg}.  The corresponded neutrino flux from the bubbles is consistent with the hadronic flux model that reproduce FB gamma ray data. If that is the case, the bubbles could be the first multi-messenger study source in our milky way \cite{Lunardini:2015laa, Fang:2017vlg}. Recently, the High Altitude Water Cherenkov (HAWC) telescope reported their analysis of gamma-ray data from the high-latitude ($b>6^\circ$) Northern FB region collected over 290 days \cite{Abeysekara:2017wzt} and found no evident excess above $\sim 1$ TeV. The HAWC upper limits are consistent with the gamma-ray flux from \textit{Fermi}-LAT but constrained our previous ``neutrino-inspired'' hadronic models that predict high gamma-ray flux level in the $\ge 1$ TeV range from the whole FB region \cite{Lunardini:2015laa}. 

The morphology of the FB gamma-ray emission, however, is not fully defined, especially toward the GC region, which has also not been clearly observed in other wavelength.  In the latest \textit{Fermi}-LAT analysis \cite{TheFermi-LAT:2017vmf}, it was found that the spectra of the low-latitude ($|b| < 10^\circ$) and high-latitude ($|b| > 10^\circ$) bubbles are similar in the 100 MeV to 100 GeV energy range, but different at energies above 100 GeV.  In particular the spectrum of the low-latitude bubbles is hard and without any apparent fall-off as compared to the high-latitude bubble spectrum which shows a sharp cutoff above $\sim 100$ GeV \cite{TheFermi-LAT:2017vmf}.  As mentioned, the HAWC upper limits \cite{Abeysekara:2017wzt} do not constrain the low-latitude FB spectrum. The HAWC upper limits derived for the North bubble also do not strictly apply to the South bubble in case the South bubble has different gamma-ray morphology and spectrum than the North bubble. The sample of the high-energy astrophysical neutrinos detected by IceCube has grown to 82 in the meantime (year 2010-2016) \cite{Aartsen:2017mau} and several more of these events are now from the FB region as compared to what we had reported in \cite{Lunardini:2013gva, Lunardini:2014wza}. Given these developments, it is relevant now to critically review and update the FB hadronic models with relevant constraints.

In this article we present updated hadronic models for the FB, based on the recent gamma-ray spectral analysis by the \textit{Fermi}-LAT Collaboration with 6.5 years data \cite{TheFermi-LAT:2017vmf}, using the high- and low-latitude bubble templates, latest sample of astrophysical neutrinos detected by IceCube and HAWC constraints at TeV energy gamma ray observations. A new feature in the latest Fermi-LAT data is that the gamma-rays from the FB at the high- and low- Galactic latitudes have different spectra. Therefore we present two hadronic models in the present work fitting those spectra. We take (in this work and in Refs. \cite{Lunardini:2013gva, Lunardini:2014wza,Lunardini:2015laa}) both the gamma-ray and neutrino data into account to constrain the primary proton spectra instead of calculating the proton distribution function fitting the gamma-ray data only (see, e.g., Ref. \cite{Fujita:2014oda}). The neutrino data to is especially useful to constrain the high-energy cutoff in the primary proton spectrum for the low-latitude case. We discuss details of the latest gamma-ray and neutrino data in Sec.\ 2, develop updated hadronic models  in Sec.\ 3 and discuss possible future constraints from gamma-ray observations in Sec.\ 4.  We summarize and conclude in Sec.\ 5.

\section{Recent gamma-ray and neutrino data}

\begin{figure}[htb]
\centering
\includegraphics[width=10 cm]{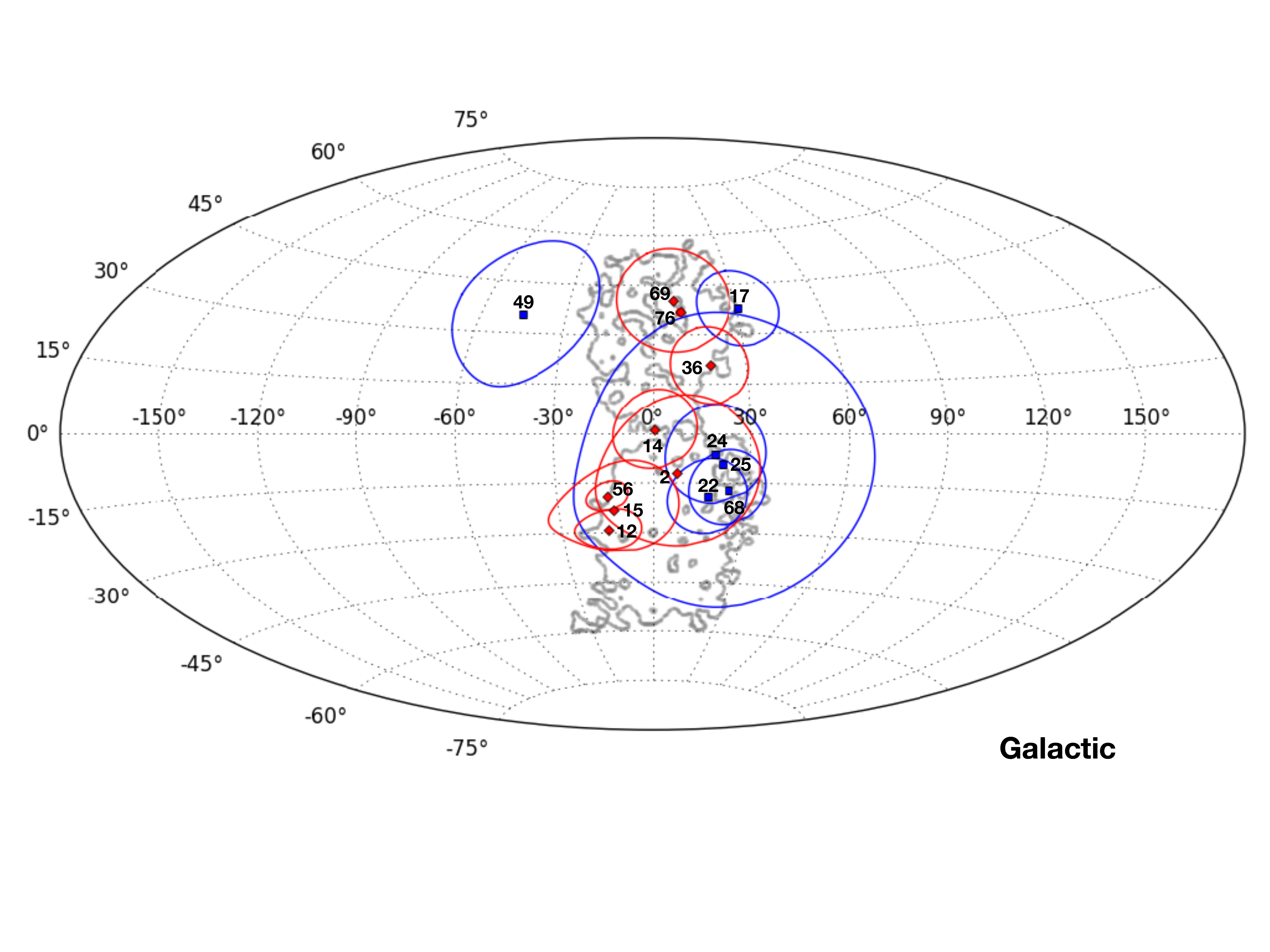}
\caption{Sky map of the IceCube astrophysical neutrino events \cite{Aartsen:2017mau} that are correlated with the \textit{Fermi} bubbles in Galactic coordinates. The eight strongly- and six weakly-correlated events are shown as red diamonds and blue squares, respectively, with their median angular errors and numbers labeled. The \textit{Fermi} bubbles outlines are shown as gray contours.}
\label{fig:icecube}
\end{figure}

The \textit{Fermi} bubbles as analyzed with \textit{Fermi}-LAT data \cite{Fermi-LAT:2014sfa, TheFermi-LAT:2017vmf}, were found to have uniform spectra above 10$^\circ$ in Galactic latitude with sharp edges. In the most recent update with 6.5 years of \textit{Fermi}-LAT data, the spectra and morphology of central bubbles have also been derived with the spectral components analysis procedure as in \cite{Fermi-LAT:2014sfa}.  In this latest analysis, the spectra of low- and high-latitude bubbles are treated separately, but with the assumption that they behave similarly in the energy between 1 GeV and 10 GeV. In addition, the center of the bubbles is found to become brighter as they are close to the Galactic plane, which is different from previous models \cite{TheFermi-LAT:2015kwa} with isotropic emission in the center region. In our study, we adopted the templates derived by \textit{Fermi} Collaboration with the region of interest |b| < 60$^\circ$,  |l| < 45$^\circ$, as shown in gray contours in Fig.~\ref{fig:icecube}, obtaining a solid angle of $\Omega_{FB}=$1.04 sr and fit our models with respect to the low- and high-latitude components separately as well.

\begin{figure}[htb]
\centering
\includegraphics[width=10 cm]{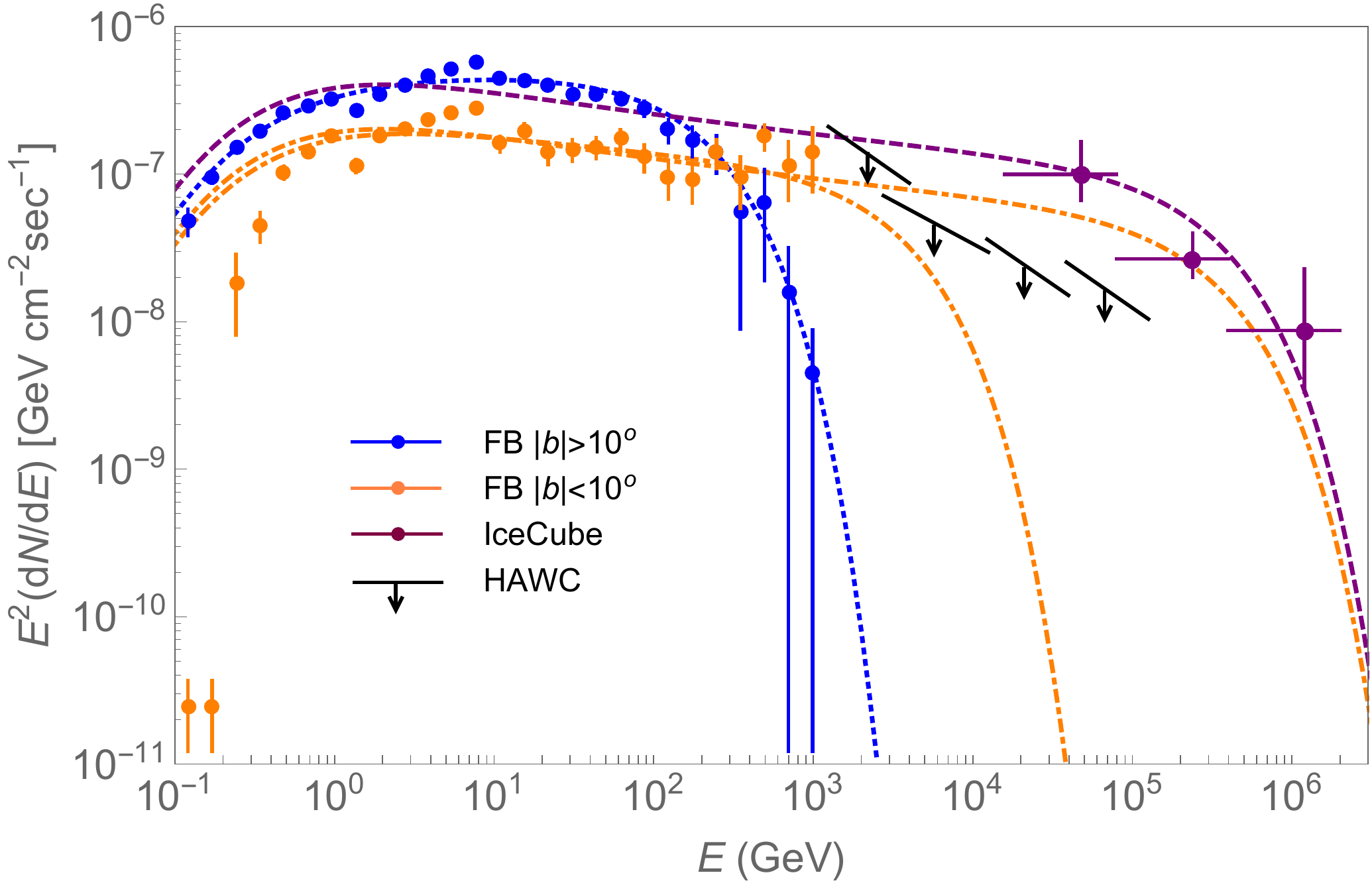}
\caption{\textit{Fermi}-LAT spectra from low- and high-latitude regions of the FB \cite{TheFermi-LAT:2017vmf}.  Also shown are HAWC upper limits in the $b>6^\circ$ region \cite{Abeysekara:2017wzt} and neutrino spectra, assumed originated from the FB, from 8 events \cite{Aartsen:2017mau}. The hadronic models with primary proton spectra $\propto E^{-k} \exp(-E/E_0)$ are shown with lines for $k=2.0$, $E_0 = 1.6$ TeV (blue dotted lines); $k=2.15$, $E_0 = 30$ TeV and $k=2.2$, $E_0 = 3$ PeV (orange dot-dashed lines). The magenta dashed line correspond to the neutrino flux (all three flavors) for the model with $k=2.2$, $E_0 = 3$ PeV.}
\label{fig:spectra}
\end{figure}

Recently, the HAWC Collaboration reported their search for very high energy (VHE) gamma-ray emission from the Northern FB region with latitude above 6$^{\circ}$ which corresponds to a FB solid angle of 0.42 sr \cite{Abeysekara:2017wzt}. No significant excess was found in the analysis and the 95\% C.L. upper limits consistent with HAWC detection power on the differential flux in four energy bins were obtained as shown in Fig.~\ref{fig:spectra}. The limits agree with gamma ray flux above $\sim 2$ TeV from the FB regions at high-latitudes.

Lately, the IceCube Neutrino Observatory has also updated its high-energy starting events (HESE) search with the neutrino interaction vertex inside the IceCube detector volume and energies above 30 TeV, using six-year dataset, which corresponds to a total livetime of 2078 days. In the full sample, there are 82 events detected, with expected atmospheric muon background of 25.2 $\pm$ 7.3 events \cite{Aartsen:2017mau}.  Of these events, eight (Nos.\ 2, 12, 14, 15, 36, 56, 69, 76) are spatially strongly correlated (the best-fit arrival direction is within the bubble geometry) and six (Nos.\ 17, 22, 24, 25, 49, 68) are weakly correlated (the median positional error circles overlap with the bubble geometry) with the FB, as seen in Fig.~\ref{fig:icecube}.  Except for the event No.\ 76, which is a track, all the other events are showers. In particular, within these eight strongly correlated neutrino events, three of them are from high-latitude northern bubble, four from high-latitude southern bubble, one with the highest energy $\sim 1$ PeV is from the Galactic Center region.  Fig.~\ref{fig:icecube} also shows the median positional error for the neutrino arrival directions.

In Fig.~\ref{fig:spectra} we show the neutrino spectrum (all three flavors) we have calculated from the 8 events in three energy bins from the FB region using the livetime and effective areas of IceCube for HESE analysis \cite{Aartsen:2017mau}. There are 3, 4 and 1 event(s), respectively, in the energy ranges ($10^{4.2}$--$10^{4.9}$) GeV, ($10^{4.9}$--$10^{5.6}$) GeV and ($10^{5.6}$--$10^{6.3}$) GeV.  The flux errors on the data points are calculated assuming Poisson statistics.

\section{Hadronic models}

Prolonged star-formation activity near the Galactic center, which forms a strong bipolar wind above and below the Galactic plane, has been proposed as a plausible mechanism to inflate the FB \cite{Crocker:2010dg}.  Alternately, activity of the Galactic super-massive black hole in the past could also be responsible for the FB origin \cite{Su:2010qj}.  In the hadronic models, the gamma rays from the FB are produced primarily from neutral pion decays which are created by proton-proton ($pp$) interactions of energetic cosmic rays with the dilute gas inside the bubble volumes.  Cosmic rays, accelerated to $\sim 10^{15}$ eV and possibly beyond in supernova remnants, can be carried by the wind to fill the FB volume.  Decays of charged pions and kaons, co-produced with neutral pions, by $pp$ interactions guarantee a neutrino flux associated with the gamma-ray flux in the hadronic models.

In our hadronic model formalism \cite{Lunardini:2011br, Lunardini:2013gva, Lunardini:2015laa} we assume a primary proton spectrum in the form of a power law with exponentially cutoff as $N_{p}(E) = N_0 E^{-k} \exp(-E/E_0)$.  Here $N_0$, $k$ and $E_0$ are the normalization factor, spectral index and cutoff energy, respectively.  The normalization factor $N_0$ is adjusted to fit the gamma-ray spectra by using a particle density of $10^{-2}$ cm$^{-3}$ for the gas inside the bubble volumes.  The spectral index and the cutoff energy are crucial parameters which can be utilized to constrain the hadronic models from multi-wavelength and multi-messenger observations.

In Fig.~\ref{fig:spectra} we show our hadronic model fits for both high-latitude (FB $|b|>10^\circ$) and low-latitude (FB $|b|<10^\circ$) spectra of the FB.  The high-latitude spectrum is rather well-fitted with $k=2.0$ and $E_0 = 1.6$ TeV (blue dotted lines).  This cutoff energy is consistent with the HAWC upper limits in the $\sim 2$--200 TeV range, which apply dominantly to the high-latitude part ($b>6^\circ$) of the North bubble \cite{Abeysekara:2017wzt}.  We show two model fits (orange dot-dashed lines) for the low-latitude spectra in Fig.~\ref{fig:spectra}, both fitting low-energy ($\le 10$ GeV) data rather poorly.  The low-energy part of the spectrum is less reliable due to uncertainties in the Galactic emission template at low latitudes and possible contamination with Galactic center emission \cite{TheFermi-LAT:2017vmf}, such as the point source detection, morphology study of bubbles and analysis of Galactic Interstellar Emission.  The model with lower cutoff energy has parameters $k=2.15$ and $E_0 = 30$ TeV and the one with higher cutoff energy has parameters $k=2.2$ and $E_0 = 3$ PeV.  All values of $k$ and $E_0$ are consistent with shock-acceleration scenario of cosmic rays in supernova remnants.

The neutrino flux (all three flavors) arising from the hadronic model with $k=2.2$ and $E_0 = 3$ PeV is shown in Fig.~\ref{fig:spectra} with magenta dashed lines.  Although such a high cutoff energy is not required by the \textit{Fermi}-LAT gamma-ray data alone, an explanation of the neutrino data could be provided with this model.  This is also motivated by the hard spectrum of the low-latitude FB without showing any cutoff, unlike the high-latitude spectrum, and that the HAWC upper limits do not strictly apply in this region.  Of course the neutrino events that are strongly correlated with the FB should mostly be arriving from the $|b|<10^\circ$ regions in this scenario, which could be plausible given large uncertainties in reconstructing their directions \cite{Aartsen:2017mau}.  Alternately, only a fraction of these 8 events could originate from the FB and the rest would be from a diffuse astrophysical background.  We elaborate on this scenario in the following subsection.

\subsection{Neutrino events from the FB}

In the recent two-year dataset, IceCube Observatory detected 28 more HESE neutrinos, and notably all of them are with energy below 200 TeV \cite{Aartsen:2017mau}. Altogether there are now 82 events detected so far and as Nos.\ 20 and 55 are in coincident with background muons, they are excluded from the analysis. The likelihood fitting was performed in the deposited energy between 60 TeV and 10 PeV by IceCube \cite{Aartsen:2017mau}, resulting a best fit single power law flux of $dN/dE = 2.46 \times(E/100\,{\rm TeV})^{-2.92}$~GeV$^{-1}$~cm$^{-2}$~s$^{-1}$~sr$^{-1}$ per flavor, which is softer than the previous results \cite{Aartsen:2013jdh, Kopper:2015vzf} due to the additional events at low energies. Initially IceCube focused on the higher-energy search around PeV \cite{Aartsen:2013jdh}, later the selection threshold of deposited energy was reduced down to $\sim 1$ TeV \cite{Aartsen:2014muf}, where cosmic ray muon background is dominant. Applying the veto technique as in \cite{Aartsen:2014muf} and also the first-level online filters, resulted in an increase in the effective area \cite{Aartsen:2017mau} for both cascade-like and track-like events.

\begin{figure}
    \centering
    \subfloat[Neutrino fluxes]{{\includegraphics[width=7.5cm]{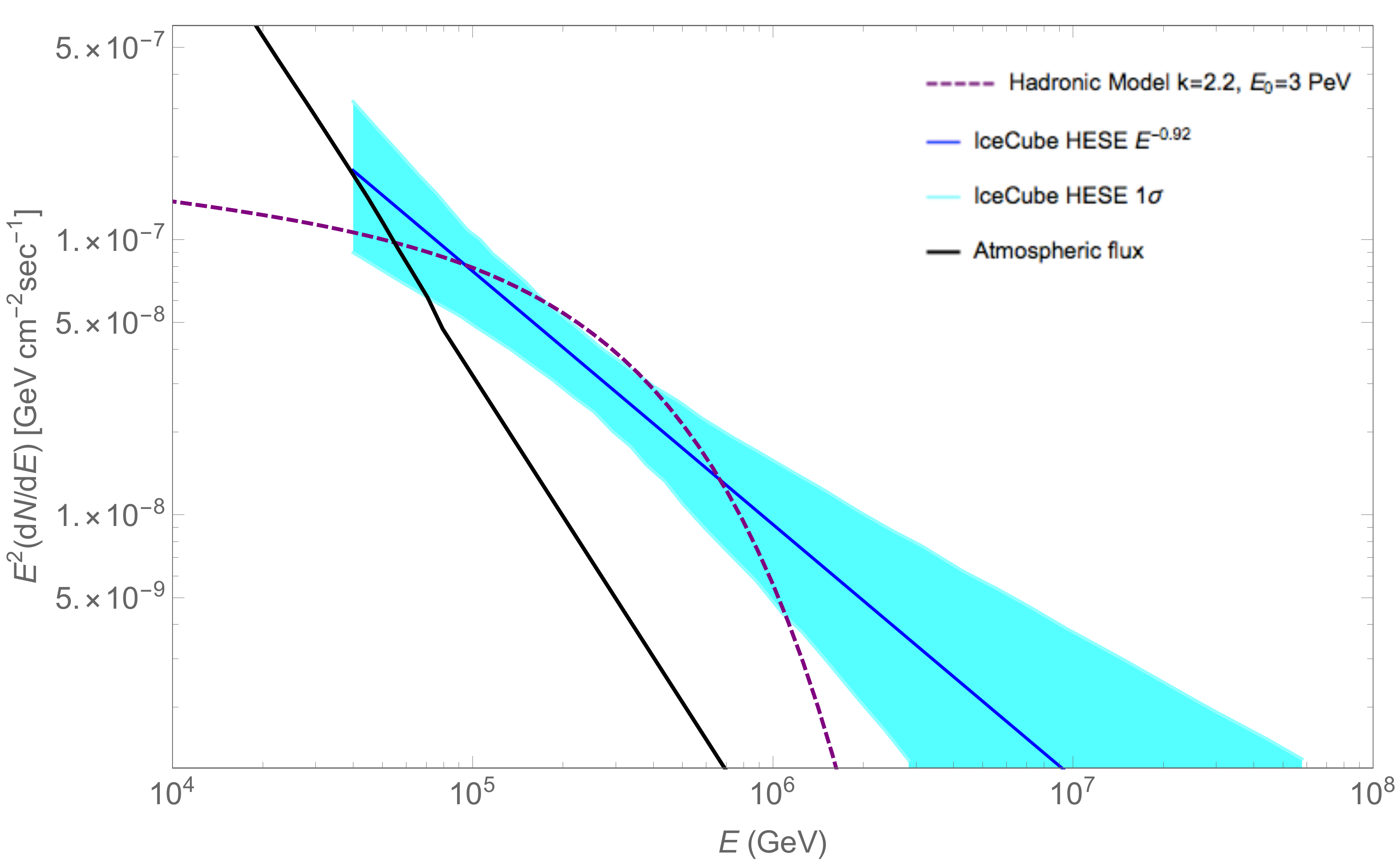} }}%
    \qquad
    \subfloat[Neutrino event rate]{{\includegraphics[width=7.cm]{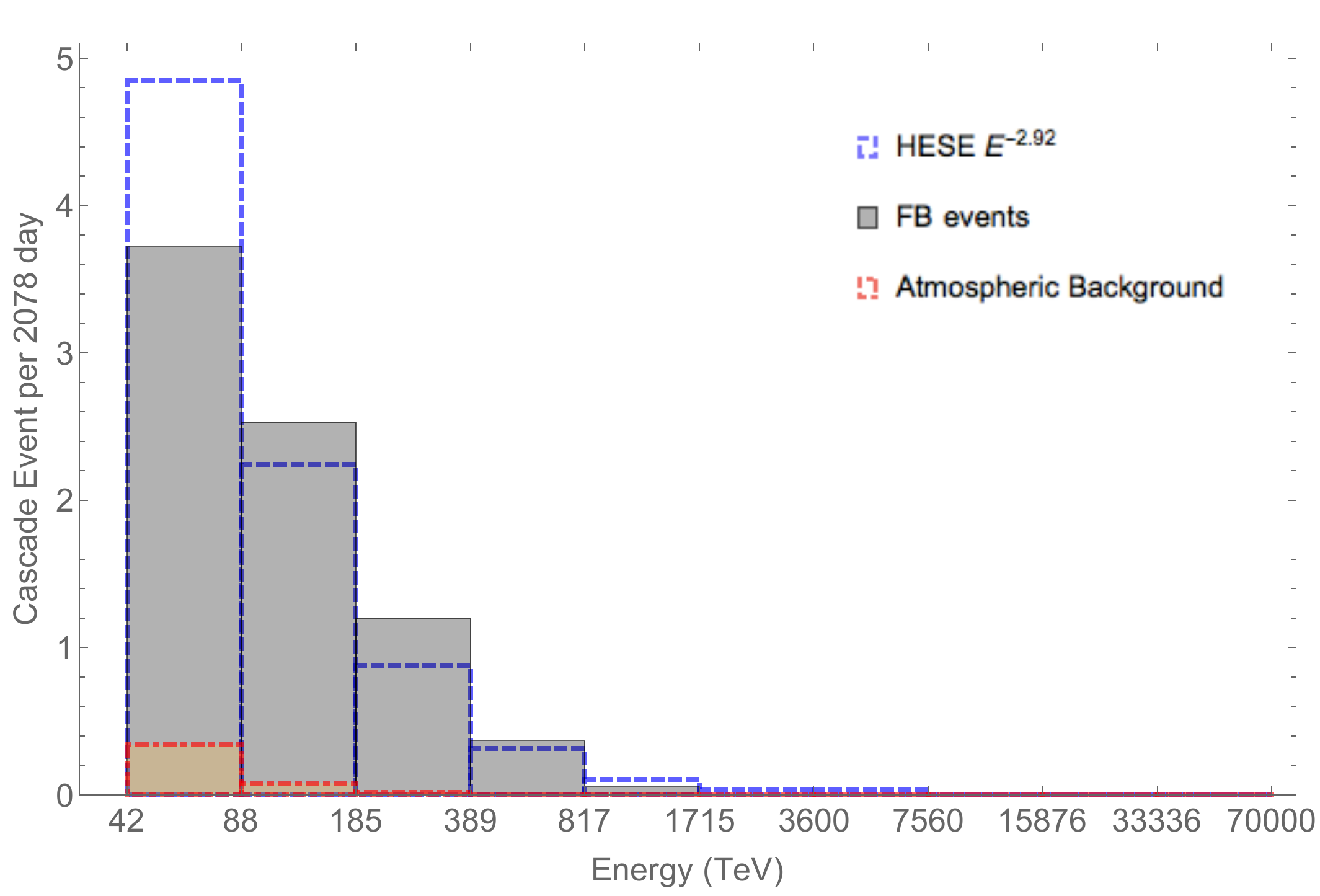} }}%
    \caption{(a) Best-fit astrophysical neutrino fluxes with a single power law of $E^{-0.92}$ and one sigma uncertainties as presented in blue shade. The black line represents the atmospheric neutrino fluxes and the magenta line is the hadronic flux in Fig. 2. (b) Expected number of neutrino events from FB, astrophysical and atmospheric backgrounds are shown in gray, red and white (dashed blue edge) bars.}%
    \label{fig:fluxandevents}%
\end{figure}

To estimate the neutrino signal excess from the FB in 2078 days operation of IceCube, the major backgrounds are atmospheric muons, neutrinos of astrophysical origin and atmospheric neutrinos in the energy range of 40 TeV to 70 PeV. The first case strongly relies on the efficiency of background rejection and reconstruction technique (see, \cite{Aartsen:2013jdh, Aartsen:2017mau} for more  information). The predicted atmospheric $\nu_\mu + \bar{\nu}_{\mu}$ flux from \cite{Honda:2006qj}, averaged over the declination angle of the FB as seen from IceCube, is adopted and extrapolated at high energy. Moreover, the atmospheric $\nu_e + \bar{\nu}_{e}$ fluxes are greatly suppressed compare to $\nu_\mu + \bar{\nu}_{\mu}$ by a factor of 14 \cite{Sinegovskaya:2013wgm}, which has also been taken into account. The astrophysical neutrino fluxes from the IceCube best fit adopted here are shown in Fig.~\ref{fig:fluxandevents}(a) as blue line with one-sigma uncertainties (shaded bands), and the hadronic fluxes from the FB region as in Fig.~\ref{fig:spectra} is shown as well. All plotted fluxes in Fig.~\ref{fig:fluxandevents}(a) are for a solid angle of FB $\Omega_{\text{FB}}$=1.04 sr for all three flavors (combined neutrino and antineutrino).

Since all the neutrinos spatially correlated with FB are cascade events, except the No.\ 76, we calculate the number of cascade events only with updated HESE effective areas as seen in Fig.~\ref{fig:fluxandevents}. The number of events in each energy bin $i$ can be calculated as $N_i=\int_{E_{\text{i, min}}}^{E_{\text{i, max}}} (dN/dE)A_{\text{eff}}T_{\text{exp}}\Omega_{\text{FB}}$,
%
%
where $A_{\text{eff}}$ and $T_{\text{exp}}$ are $\nu_e$ or $\nu_\tau$ effective area and exposure time (2078 days), respectively.  These event distributions are plotted in Fig.~\ref{fig:fluxandevents}.  We expect a total of 8.8 and 7.9 cascade-like events from the backgrounds (atmospheric + astrophysical) and FB, respectively, in the first five energy bins. Given the harder spectrum of the FB flux than the diffuse astrophysical flux, the signature of the FB is more prominent in the $\sim 100$ TeV--1 PeV range. Within one sigma uncertainty, the number of background events can be reduced to 5.5. The estimation agrees with the IceCube observation, where 8 (6) events are strongly (weakly) correlated with the FB.  One note of caution, however, is that the diffuse flux fit is based on all events, including plausible contribution from the FB.

Detection of track-like neutrino events, which have $\le 1^\circ$ angular resolution, by the upcoming KM3NeT Neutrino Telescope located in the northern hemisphere \cite{Adrian-Martinez:2016fdl} will be helpful to separate the FB from other possible Galactic sources, as well as complimentary to the IceCube data dominated by shower-like events.

\section{Future VHE gamma-ray constraints on hadronic models}

The HAWC observatory found no significant excess towards the FB using the data between 2014 November 27th to 2016 February 11th, resulting in a 90\% upper limits on high-latitude northern bubble with b > 6$^\circ$ in the energy between 1 and 100 TeV \cite{Abeysekara:2017wzt}.  These upper limits agree with our hadronic emission model of the high-latitude FB, which is strongly suppressed beyond $\sim 1$ TeV (see Fig.~\ref{fig:spectra}).  The spectrum of the low-latitude FB is harder and without any apparent cutoff in the energy range of the \textit{Fermi}-LAT.  Therefore both our nominal model with  parameters $k=2.15$ and $E_0 = 30$ TeV and the neutrino-inspired model with parameters $k=2.2$ and $E_0 = 3$ PeV for the low-latitude spectrum are potential targets for future VHE observations.  We expect that the HAWC collaboration will further analyze data from the FB in future, including the sensitivity at lower energies and larger field of view that includes the Galactic Center region.  This will provide more stringent constraints on both the spectra at low- and high-latitude bubbles, the full image of non-uniform intensity and the shape of the central region.

\begin{figure}[H]
\centering
\includegraphics[width=10 cm]{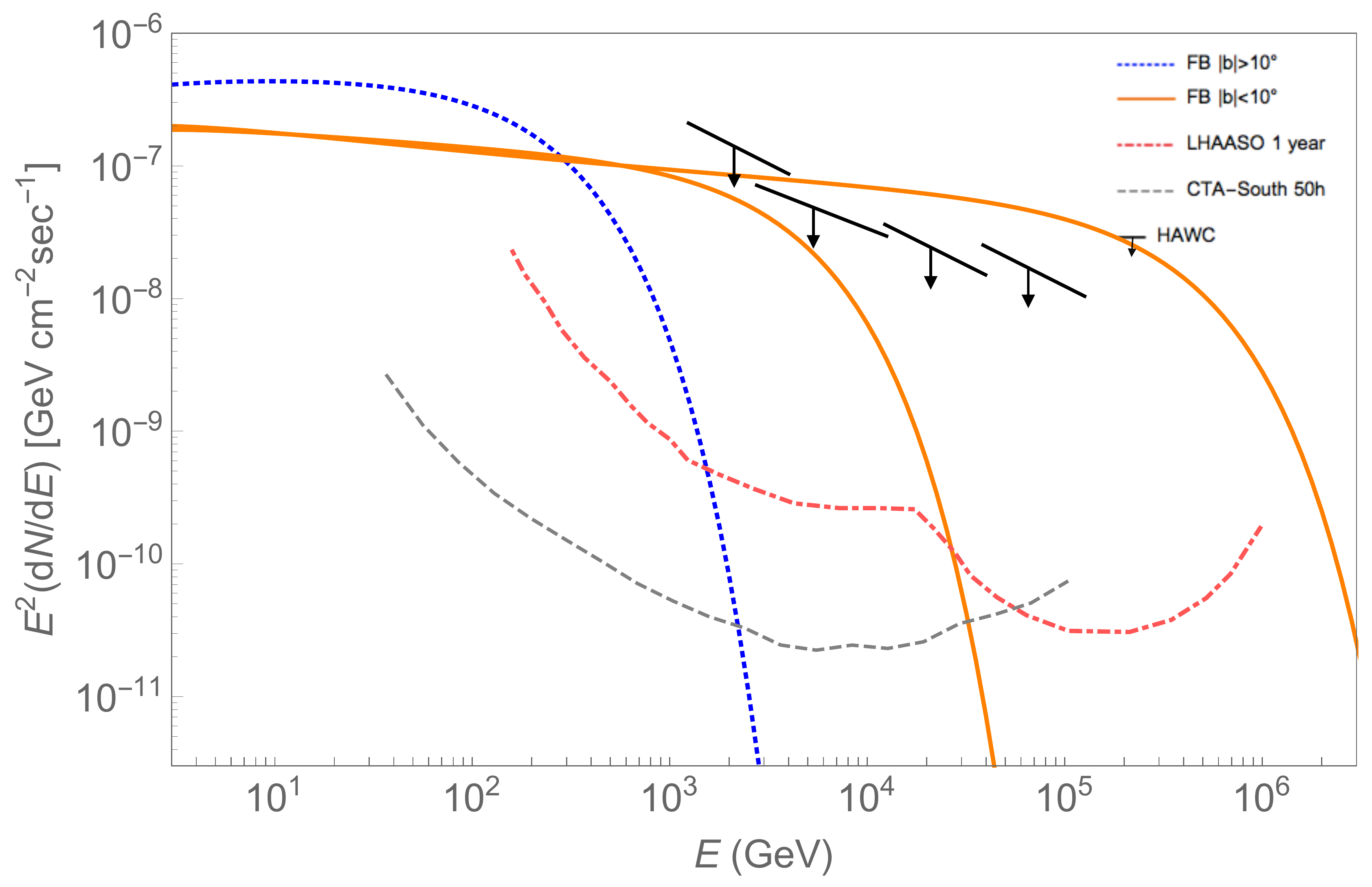}
\caption{FB spectra from the low- and high-latitude regions and HAWC upper limits as shown in Fig.~\ref{fig:spectra}. The differential point source flux sensitivities of the CTA southern site \cite{Acharya:2017ttl} with 50 hour observation (gray dashed curve) and of the LHAASO \cite{DiSciascio:2016rgi} with 1 year observation (red dot-dashed curve) are shown with the FB flux models for illustration purpose.}
\label{fig:sens}
\end{figure}

On the other hand, there is good hope that the Cherenkov Telescope Array, the next generation of ground-based VHE observatory \cite{Acharya:2017ttl} with a northern (La Palma) and a southern (Chile) site, will be able to observe both bubbles. A detailed strategy needs to be developed about pointing different parts of the FB as well as better controlling the background. In Fig.~\ref{fig:sens} we show the differential sensitivity of the CTA-South for 50 hour observation of a point source \cite{Acharya:2017ttl}, for illustration purpose.  Moreover, the Large High Altitude Air Shower Observatory (LHAASO) being built at 4410 m altitude in Sichuan Province of China, detecting cosmic rays and gamma rays in the energy range of 100 GeV to 1 PeV \cite{DiSciascio:2016rgi}, will be complementary to the HAWC and CTA. With a high duty cycle and wide filed of view, $\approx \pi$ sr, LHAASO will be able to observe the full northern bubble with a solid angle of 0.45 sr.  We show the differential sensitivity of LHAASO in Fig.~\ref{fig:sens} for one year observation of a Crab-like point source \cite{DiSciascio:2016rgi}, again for illustration purpose. A detailed spectral and morphological study with simulations will be required to calculate the sensitivities of the CTA and HAWC to the FB.


\section{Summary and Outlook}
In this paper we have carried out updated hadronic modeling of the  0.1 GeV -- 1 TeV gamma-ray emission from the FB, based on new data from the \textit{Fermi}-LAT, upper limits from the HAWC and expanded astrophysical neutrino sample from the IceCube.  Latest analysis of \textit{Fermi}-LAT data prefers a spectrum with cutoff above 1 TeV in the high Galactic latitude ($|b| > 10^\circ$) region of the FB while a hard spectrum without any apparent cutoff in the low Galactic latitude ($|b| < 10^\circ$) region of the FB  \cite{TheFermi-LAT:2017vmf}.  Combined with the latest IceCube astrophysical neutrino data \cite{Aartsen:2017mau} our updated hadronic model can be extended to the 1 PeV range for the low-latitude FB emission and is not directly constrained by the HAWC upper limits derived from observations of the northern bubble at high-latitude ($b>6^\circ$) \cite{Abeysekara:2017wzt}.  Our updated hadronic model for the high-latitude FB spectrum is consistent with the HAWC upper limits.

The observation of high-energy neutrinos from the FB would strongly support the hadronic origin of gamma rays.  Detection of several more astrophysical neutrino events by IceCube from the direction of the FB in the latest dataset is therefore intriguing.  Our estimated FB neutrino flux based on these events is harder than a diffuse astrophysical flux estimate by the IceCube collaboration, although it is not possible to distinguish them using the current data set.  Future observation of neutrinos by KM3NeT \cite{Adrian-Martinez:2016fdl} from the FB region can shed insights, especially if many track-like events are detected with good angular resolution.  Observations of VHE gamma rays from the FB with upcoming CTA \cite{Acharya:2017ttl} and LHAASO \cite{DiSciascio:2016rgi}, combined with deeper observations by HAWC, will be crucial to probe the spectra of FB in the 1--100 TeV range that ties with the lower energy range of the IceCube neutrino events.  The detection of a cutoff in this energy range for the low-latitude FB spectrum could strongly constrain hadronic models that can explain the IceCube neutrino data and critically test the FB as a multi-messenger source.

\vspace{6pt}

\acknowledgments{We thank D.\ Malyshev for providing latest \textit{Fermi}-LAT spectral data.  S.R.\ acknowledges support from the National Research Foundation (South Africa) with Grant No: 111749 (CPRR).  L.Y.\ acknowledges the Slovenia Research Agency grant number Z1-8139 (B).}

\authorcontributions{S.R.\ and L.Y.\  conceived and designed the investigation; both contributed equally to analyze the data; to perform modeling; and to write the paper.}

\conflictsofinterest{The authors declare no conflict of interest. The funding sponsors had no role in the design of the study; in the collection, analyses, or interpretation of data; in the writing of the manuscript, and in the decision to publish the results.}


\reftitle{References}

\end{document}